\documentclass[aps,showpacs,floatfix]{revtex4}
\usepackage{graphicx}
\newcommand{\be}{\begin{eqnarray}}
\newcommand{\ee}{\end{eqnarray}}
\def\beq{\begin{equation}}
\def\eeq{\end{equation}}

\begin{document}
\baselineskip18pt
\title{Anderson localization in quantum chaos: scaling and universality}
\author{Antonio M. Garc\'{\i}a-Garc\'{\i}a}
\affiliation{Physics Department, Princeton University, Princeton,
New Jersey 08544, USA,} \affiliation{The Abdus Salam International
Centre for Theoretical Physics, P.O.B. 586, 34100 Trieste, Italy}
\author{Jiao Wang}
\affiliation{Temasek Laboratories, National University of
Singapore, 117542 Singapore,} \affiliation{Beijing-Hong
Kong-Singapore Joint Center for Nonlinear and Complex Systems
(Singapore), National University of Singapore, 117542 Singapore}

\begin{abstract}
The one parameter scaling theory is a powerful tool to investigate
Anderson localization effects in disordered systems. 
In this paper we show this theory can be adapted to the context of quantum
chaos provided that the classical phase space is homogeneous, not
mixed.  The localization problem in this case is defined in momentum, not in real space. 
We then employ the one parameter scaling theory to: a) propose a
precise characterization of the type of classical dynamics related
to the Wigner-Dyson and Poisson statistics which also predicts in
what situations Anderson localization corrections invalidate the
relation between classical chaos and random matrix theory encoded
in the Bohigas-Schmit-Giannoni conjecture, b) to identify the universality class associated with the
metal-insulator transition in quantum chaos.  In low dimensions it
is characterized by classical superdiffusion, in higher dimensions
it has in general a quantum origin as in the case of disordered
systems.  We illustrate these two cases by studying 1$d$ kicked
rotors with non analytical potentials and a 3$d$ kicked rotor with
a smooth potential.
\end{abstract}

\maketitle

The Bohigas-Giannoni-Schmit (BGS) conjecture \cite{oriol} states
that quantum spectra of classically chaotic systems are
universally correlated according to the Wigner-Dyson random matrix
ensembles (WD).  Therefore for scales of time comparable to the
Heisenberg time the quantum system evolves to a state where all
dynamical features are washed out,  and all the available phase
space is occupied. Eigenstates are fully delocalized  and the
spectral correlations are dictated by symmetry only.
For the sake of completeness we mention the relation between level repulsion and classical chaos
 was noted previously in  
\cite{kauf,casati}.  However we stick to the term "BGS conjecture" through the paper since O. Bohigas and coworkers were the first
to claim that a) this relation was universal, b) spectral correlations are
described by random matrix theory. From our point of view this is a qualitative difference from the findings of Ref.\cite{kauf,casati}.

Similarly, according to the Berry-Tabor-Gutzwiller (BTG)
conjecture \cite{tabor}, the spectral correlations of classically
integrable systems are well described by Poisson statistics.
According to the Noether theorem there is a conserved quantity
(generalized momentum) for each classical symmetry. Integrability
assures that the number of these conserved momenta is equal to the
dimensionality of the space. In quantum mechanics each conserved
quantity leads to a good quantum number. A given quantum state is
thus characterized by a unique set of good momentum quantum
numbers. Since different states are characterized by a different
set of quantum numbers, the spectrum is not correlated. In other
words, in integrable
systems the Hamiltonian is diagonal in a basis of momentum eigenstates. \\

%By modifying the parameters of the system, a transition from integrable to chaotic dynamics can be observed. If
%the KAM theorem is applicable, this transition is smooth and both integrable and chaotic
%regions coexist until the last KAM torus is completely destroyed.
%In this intermediate situation the spectral correlations are non universal. The time evolution of a wave packet strongly depends thus on details such as
%initial conditions.  Neither in the mixed nor in the chaotic case
%it is possible to find a basis in which the Hamiltonian is diagonal, therefore the spectrum, at least in some degree,  is correlated .

The influence of these two conjectures in the establishment and
development of quantum chaos is hard to overestimate.  The reason
is simple: Comparatively it is straightforward to determine
whether the classical dynamics is chaotic or integrable. With this
input one can predict the quantum transport properties of the
system even without a full consideration of the specific
details of the Hamiltonian. \\

However, exceptions are well documented in the literature,  among
others we cite arithmetic billiards \cite{ari}, Harper model
\cite{harper}, and the ubiquitous kicked rotor \cite{kick},
namely, a free particle in a circle which experiences periodic
kicks of amplitude given by a periodic smooth potential, \be {\cal
H}= \frac{p^2}{2} - K\cos(q)\sum_{n}\delta(t-Tn).\label{kr} \ee
For short time scales, both quantum and classical motion is
diffusive in momentum space. However, quantum diffusion is
eventually arrested due to destructive interference
effects that localize eigenstates in momentum space. In this limit
spectral correlations are described by Poisson but not by WD
statistics. This counter-intuitive feature, usually referred to as
dynamical localization \cite{kick}, was fully understood
\cite{fishman} after mapping the kicked rotor problem onto a short
range one dimensional disordered system where localization is well
established.  \\

Deviations from the BGS conjecture are also expected \cite{larkin}
for eigenvalue separations $\delta E \sim \hbar/t_E$ due to weak
localization effects. The typical scale  $t_E = \lambda^{-1}|\log
\hbar|$ is the Ehrenfest time with $\lambda$ the classical
Lyapunov exponent. These corrections do not invalidate the BGS
conjecture since $\delta E \sim \hbar/t_E$ is much larger than the
mean lever spacing $\Delta$, the natural scale in which WD statistics holds.  \\

Another potential source of  deviations from WD statistics is
related to energy scales larger than $\hbar / t_{fl}$ where
$t_{fl}$ is the typical time for the classical dynamic to become
fully chaotic. In billiards it is just the flight time of a
particle across the system. These corrections can be expressed
analytically \cite{andre,andre1} in terms of the spectral
determinant of the Perron-Frobenious operator which controls the
relaxation of the classical density to the ergodic-universal
limit described by random matrix theory. As for weak localization
corrections, the associated energy scale is much larger than the
mean level spacing and consequently does not really affect the
validity of the BGS conjecture.\\

From the previous discussion it is clear that the main deviation
from the semiclassical picture contained in the BGS conjecture is
due to strong quantum localization effects, or as it is usually
referred in the literature, Anderson localization.  After this
introduction some questions arise: Why in some quantum chaotic
systems localization is only weak and in others is strong enough
to make the system and insulator invalidating thus the BGS
conjecture? Are chaotic systems the only ones in which the BGS
conjecture applies?  If not so, is it possible to give a precise
relation between classical motion and quantum features? And is it
possible to define any other universality class in the context of
non random Hamiltonians?\\

In this paper we aim to address these questions by using ideas and
techniques originally developed in the theory of disordered systems
\cite{anderson} such as the one parameter scaling theory \cite{one}.\\

We start with a brief review that highlights some of the main ideas
and results of this field. One the main problems in the theory of
disordered systems is the study of the role of quantum effects such as
tunneling and interference on the transport properties of a
particle in a random potential. In general,  transport
properties are strongly affected by both the dimensionality of
the space ,$d$, and the strength of disorder. \\

In two and lower dimensions destructive interference caused by
backscattering produces exponential localization of the
eigenstates in real space no matter the amount of disordered
considered \cite{anderson}. As a consequence, quantum transport is
suppressed, the spectrum is uncorrelated (Poisson) and the system
becomes an insulator. In three and higher dimensions there exists
a metal insulator transition for a critical amount of disorder.
Thus for disorder below the critical one the wavefunctions are
extended through the sample, namely, they are effectively
represented by a superposition  of plane waves with random phases.
The Hamiltonian is accurately approximated by a random matrix with
the appropriate symmetry, and the spectral correlations are given
by WD statistics \cite{mehta}.  In the opposite limit, wave
functions are exponentially localized as in $d \leq 2$, and Poisson
statistics applies.\\

As in the quantum chaos case  these universal features are
restricted to long time scales (related to energy scales of the
order of the mean level spacing) such that an initially localized
wave-packet has already explored the whole phase space available.
In other words, universality is related to certain ergodic limit
of the quantum dynamics. For shorter time scales the system has
not yet relaxed to the ergodic limit and deviations from universality
are expected.\\

These features can be understood within the framework of the one
parameter scaling theory \cite{one}. A key concept in this theory
is the dimensionless conductance $g$ introduced  by Thouless
\cite{thouless2}. It is defined either as the sensitivity of a
given quantum spectrum to a change of boundary conditions in units
of the mean level spacing $\Delta \propto 1/L^d$ or as $g =
E_c/\Delta$ where, $E_c$, the Thouless energy, is an energy scale
related to the diffusion time to cross the sample.  In the
semiclassical limit $E_c = \hbar D_{cla}/L^2$ ($D_{cla}$ is the
classical diffusion constant) and therefore $g \propto L^{d-2}$.
On the other hand if the the particle is exponentially localized,
$g \propto e^{-L/\xi}$ where $\xi$ is the localization length and
$L$ is the system linear size.

The change of the dimensionless conductance with the system size
is an indicator of localization. Qualitatively, in the $L \to
\infty$ limit,  $g \to \infty$ in a metal and  $g \to 0$ in an
insulator. In order to proceed it is useful to define \be \beta(g)
= \frac{\partial \log g(L)}{\partial \log L} \ee which describes
the running of $g$ with the system size.

From the above definitions, $\beta(g) = d-2 > 0$ in a metal
(without quantum corrections) and $\beta(g)= \log(g) <0$ in an
insulator.

The one parameter scaling is based on the following two simple assumptions:

\begin{itemize}
\item The $\beta(g)$ function is continuous and monotonous; \item
The change in the conductance with the system size only depends on
the conductance itself which is a function of the system size.
\end{itemize}

We are now ready to analyze how localization corrections  depend
of the spatial dimensionality $d$:

\begin{enumerate}

\item  {\it d =1}. For $g \to \infty$, $\beta(g) = -1 < 0$.  In
the opposite limit $g \to 0$, $\beta(g) = \log (g) < 0$. Since
$\beta(g) < 0 $ for any $g$, the dimensionless conductance always
decreases with the system size and the system will be an insulator
in the $L \to \infty$ no matter the amount of disorder. \item {\it
d =3}.   For $g \to \infty$, $\beta(g) = 1 > 0$.  In the opposite
limit however $g \to 0$, $\beta(g) = \log (g) < 0$. Since
$\beta(g)$ is continuous and monotonous there must be $g =g_c$
such that $\beta(g_c) = 0$. At $g =g_c$ the system undergoes a
metal-insulator transition. The one parameter scaling theory thus
predicts a new window of universality characterized by a scale
invariant dimensionless conductance $g =g_c$. For a fixed
disorder, if $g(L) > (<) g_c$ then the system will flow to the
metallic (insulator) limit as $L \to \infty$. A similar picture
holds for $d >3$. The spectral correlations at the Anderson
transition, usually referred to as critical statistics
\cite{sko,kravtsov}, are scale invariant and intermediate between
the prediction for a metal and for an insulator \cite{sko,chi}. By
scale invariant we mean any spectral correlator utilized to
describe the spectral properties of the disordered Hamiltonian
does not depend on the system size. Eigenfunctions at the Anderson
transition are multifractals \cite{ever,schreiber} (for a review
see \cite{mirlin,cuevas}), namely, their moments present an
anomalous scaling, ${\cal P}_q=\int d^dr |\psi({\bf r})|^{2q}
\propto L^{-D_q(q-1)}$ with respect to the sample size $L$, where
$D_q$ is a set of exponents describing the Anderson transition.

\item {\it d =2}.  For $g \to \infty$, $\beta(g) =  0$ but  for $g \to
0$, $\beta(g) = \log (g) < 0$. In this case, since $\beta(g) = 0$,
the semiclassical analysis is not enough to determine the
localization properties in the thermodynamic limit.   We have to
determine whether $\beta(g)$ is positive or negative (or it
remains zero) once quantum interference effects (weak localization
corrections) are included. Technically one uses standard
perturbation theory to compute  one loop quantum corrections to
the classical diffusion constant included in the definition of the
dimensionless conductance.  The result of this calculation is that
$\beta(g) \sim -a/g$ ($a > 0$) for $g \gg 1$. Therefore $\beta(g)
< 0$ for all $g$ and, as in the $1d$ case, all eigenstates are
localized in the $L \to \infty$ for any amount of disorder. We
note that the negative sign of the weak localization correction is
not universal, it depends on the details of the Hamiltonian. We
shall see that if the classical diffusion is anomalous, weak
localization corrections renormalize the value of the classical
diffusion constant, but the semiclassical critical point $\beta(g)
=0$ is not modified.
\end{enumerate}

A few comments are in order:
\begin{enumerate}

\item The Anderson transition in a disordered conductor is a
consequence of a highly non trivial interplay  between quantum
destructive interference effects and quantum tunneling. In low
dimensions, $d \sim 2$, weak quantum destructive interference
effects induce the Anderson transition. Analytical results are
available based on perturbation theory around the metallic state
\cite{voller,wegner}. In high dimensions, $d \gg 2$, quantum
tunneling is dominant and the locator expansion \cite{anderson} or
the the self-consistent formalism \cite{abu} can be utilized to
describe the transition. We note that in these papers corrections
due to interference of different paths are neglected.

\item The above analysis is strictly valid only for time reversal
invariant systems with no interactions. A strong magnetic field or
a spin-orbit interaction may induce a metal insulator transition
in two dimensions.
\end{enumerate}

To summarize: According to the BGS conjecture, the level
statistics of quantum systems whose classical counterpart is
chaotic are described by WD statistics and eigenfunction are
delocalized in momentum space. According to the BTG conjecture,
the level statistics of quantum systems whose classical
counterpart is integrable are described by Poisson statistics and
the eigenstates are localized due to symmetry constraints. Strong
localization effects typically invalidate the relation between
classical chaos and WD statistics.

In the context of disordered systems the one parameter scaling
theory provides a valuable  framework to understand localization
effects. It predicts  three universality class: For $g \to
\infty$ the system is a metal, eigenfunctions are delocalized in
real space and the spectrum is correlated according to WD
statistics. For $g \to 0$ the system is an insulator,
eigenfunctions are exponentially localized and  the spectrum is
correlated according to Poisson statistics. For $g = g_c$ a metal
insulator-transition takes places.  The eigenstates are
multifractal and the spectral correlations are universal (critical
statistics)  but  different from WD and Poisson statistics.

The organization of the paper is as follows:  In section $2$
we discuss under what conditions the one parameter scaling theory
can be applied to quantum chaos. Then we identify a variety of
problems in quantum chaos which the one parameter scaling theory
may help solve. In section $3$  we tackle one of them: the universality class
related to the Anderson transition in quantum chaos. We test numerically the predictions of the one parameter scaling
theory about the Anderson transition in quantum chaos in a 3$d$
kicked rotor with a smooth potentials and in a  1$d$  kicked rotor
with singularities. Section $4$ is devoted to conclusions.
%%%%%%%%%%%%%Fig 1%%%%%%%%%%%%%%%%%%%%%%%%%%%%%%%%%%%%%%%%%%%%%%%%%%%%%%%%%%%%%%%%%%%%
\begin{figure*}
\includegraphics[width=.45\columnwidth,clip]{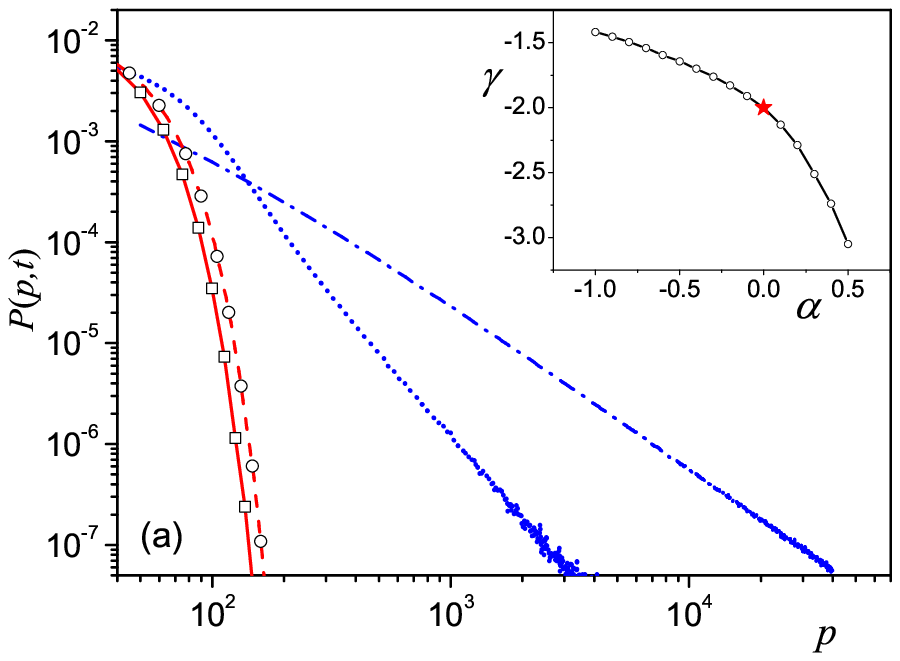}
\includegraphics[width=.45\columnwidth,clip]{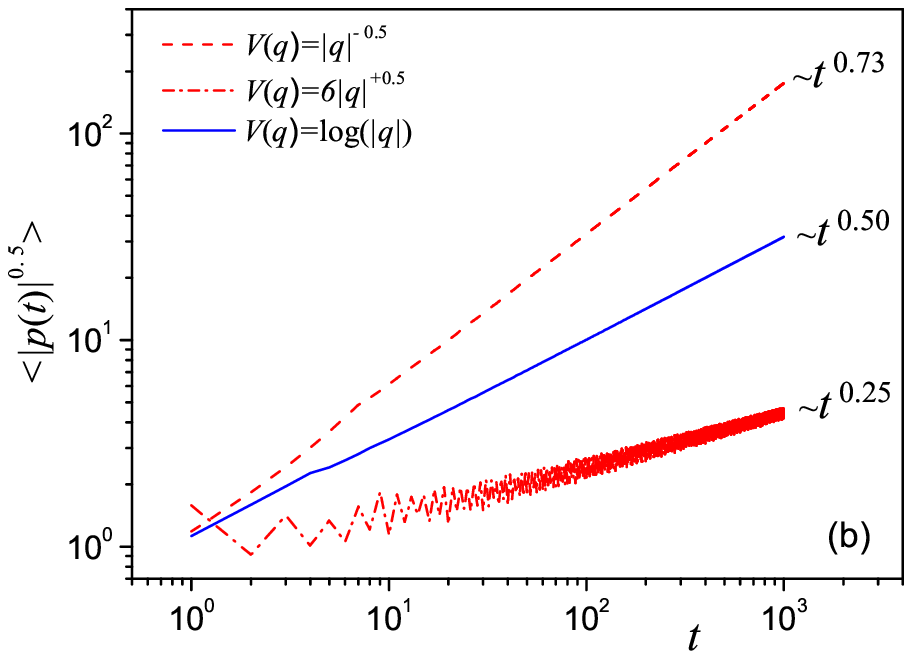}
\vspace{-.7cm} \caption{(Left)$P(p,t)$ versus $p$ ($\log$ scale).
In all cases $t=2000$ except for $\alpha=-0.5$ where $t=40$. For
$V(q)=|q|^{\alpha}$ with $\alpha=0.9$ (solid) and
$V(q)=(|q|+0.005)^{0.4}$ (dash) the diffusion is normal (squares
and circles are for best Gaussian fitting respectively). However
for $\alpha=0.4$ (dot) and $\alpha=-0.5$ (dash-dot),
$P(p,t)\propto t p^{\gamma(\alpha)}$. Inset shows $\gamma$ as a
function of $\alpha$. The star corresponds to $V(q)=\log(|q|) $.
(Right) Moments of the classical distribution for different
$\alpha's$.  As $\alpha$ increases classical diffusion is slower.} \label{figure1}
\end{figure*}
%%%%%%%%%%%%%Fig 1%%%%%%%%%%%%%%%%%%%%%%%%%%%%%%%%%%%%%%%%%%%%%%%%%%%%%%%%%%%%%%%%%%%%

%%%%%%%%%%%%%Fig 2%%%%%%%%%%%%%%%%%%%%%%%%%%%%%%%%%%%%%%%%%%%%%%%%%%%%%%%%%%%%%%%%%%%%
\begin{figure*}
\includegraphics[width=.45\columnwidth,clip]{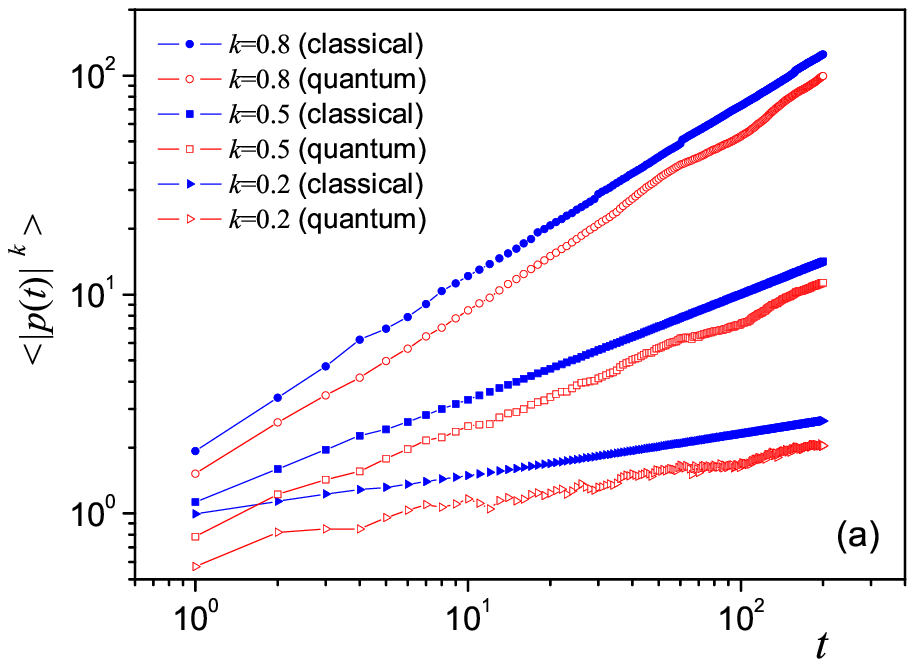}
\includegraphics[width=.45\columnwidth,clip]{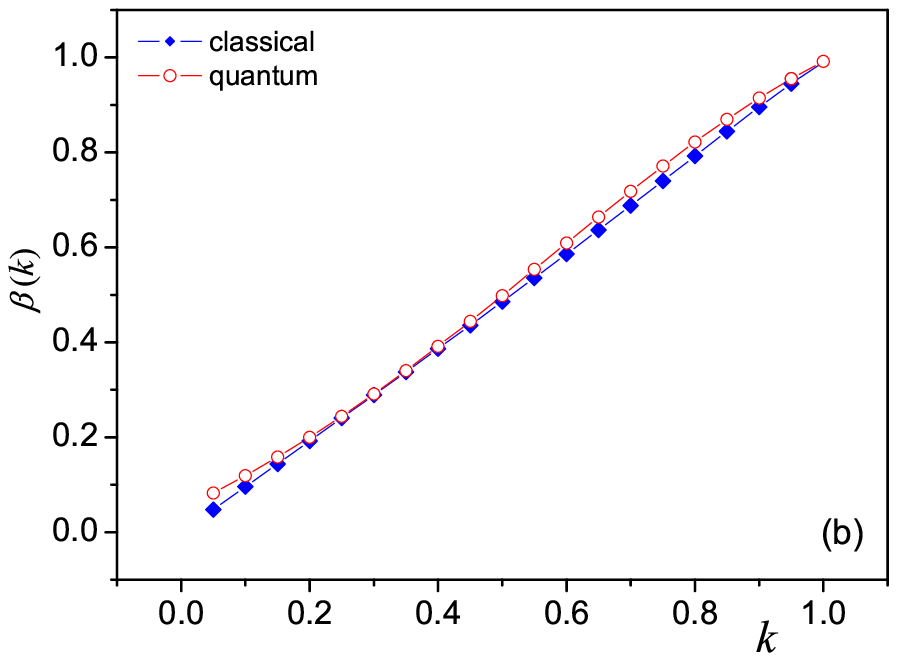}
\vspace{-.7cm}
\caption{Comparison of the classical and quantum
time dependence of the moments $\langle |p|^{k} \rangle \propto
t^{\beta(k)}$ for the case of a logarithmic singularity ($\alpha
=0$) related to the metal insulator transition.  The classical
time dependence is not modified by quantum corrections (right).  However
the value of the diffusion constant is smaller in the quantum
case (left).  This is an indication that quantum interference effects
slow down the classical motion but cannot induce full dynamical
localization.  The same conclusions apply for $\alpha < 0$.  } \label{figure2}
\end{figure*}
%%%%%%%%%%%%%Fig 2%%%%%%%%%%%%%%%%%%%%%%%%
%%%%%%%%%%%%%Fig 3%%%%%%%%%%%%%%%%%%%%%%%%%%%%%%%%%%%%%%%%%%%%%%%%%%%%%%%%%%%%%%%%%%%%

\section{One parameter scaling theory in quantum chaos}
In this section we explain how  to adapt the one parameter scaling
theory to the context of quantum chaos. The absence of a true
ensemble average in quantum chaos is in principle a major obstacle
for a straightforward application of techniques and ideas
originally developed in the field of disordered systems.
Fortunately the one parameter scaling theory is an exception. The
basic assumptions (continuity and monotonicity of the $\beta(g)$
function and its sole dependence on $g$) do not lie on any
ensemble average. Moreover the mean level spacing in the
definition of the dimensionless conductance have a well defined
meaning without having to perform an ensemble average.   The
Thouless energy, which is related to the classical diffusion time
through the sample, is typically understood as an average over
many realizations.  However this is not necessary. We just need to
find a way to determine a typical and meaningful travel time
through the sample. For instance, in kicked system it is the time
it takes to explore a momentum basis of a given size. The main
problem in quantum chaos is that in mixed systems it will depend
strongly on the initial conditions. If the particle is initially
located in a integrable region the typical time will be larger
than in a chaotic region since KAM barriers slowdown transport in
momentum space. Thus in order to define properly the dimensional
conductance in quantum chaos we have to impose the restriction
that the classical phase space must be homogeneous.
In this way a travel time can be defined unambiguously.\\

%For integrable systems transport is restricted
%by symmetry so the travel time is,  neglecting tunneling effects,
% infinite which leads to a vanishing dimensionless conductance typical of an insulator.\\

Another issue that requires clarification is the space in which
the localization problem is defined. In disordered systems is
always the real space.  Typically one fixes the strength of the
disordered potential (for instance the density of scatterers) and
then study how the dimensional conductance vary with the system
size. By contrast,  in quantum chaos the localization problem is
typically defined in momentum space. Therefore the Thouless time
is defined as the time that a wavepacket needs to explore a finite
momentum basis of size $N$ (equivalently to the system size $L$ in
disordered systems).  As in disordered systems this can be
estimated by computing the second moment $\langle p^2 \rangle$
of the distribution. \\

The reason why localization is naturally defined in momentum space
is related with the Noether theorem in classical mechanics which
states that for each  classical symmetry there is a conserved
quantity. If a system is classically integrable the number of
conserved quantities (canonical momenta)  is equal to the
dimensionality of the system. Semiclassically it is evident that
each of these canonical momenta becomes a  good quantum number
which labels the state. An integrable system is localized in
momentum space since there exists a basis of good (momentum)
quantum numbers in which the Hamiltonian is diagonal and
consequently the spectral correlations are Poisson-like as for an
insulator. If the number of conserved momenta is less than the
dimensionality of the space, the matrix representation of the
Hamiltonian in any basis of momenta  become non-diagonal and the
spectrum is correlated to some degree.  According to the BGS
conjecture, in the case of no symmetry the spectral correlations
are sufficiently strong that WD statistics applies. We note that
this semiclassical picture does not take into account of strong
localization effects that tend again to drive the system to the
insulator limit. \\

To summarize: in quantum chaos the dimensionless conductance is a
meaningful quantity provided that classical phase space is homogeneous.
The localization problem in quantum chaos is naturally defined in
momentum space. \\

The classical dynamics in quantum chaos is by no means restricted
to  standard diffusion $\langle p^2 \rangle =Dt$. Different types
of anomalous diffusion, \be \label{p2} \langle p^2 \rangle \sim
t^\beta \ee with $\beta \neq 1$ may be relevant.  
The mean level
spacing is in many cases given by $\Delta \sim 1/N^{d}$, but there
are important exceptions: i) for periodic potentials, the Bloch
theorem applies, the spectrum is continuous, and $\Delta = 0$; ii)
for systems whose eigenstates are exponentially localized, $\Delta
\neq 0$ even in the $N \to \infty$ limit; iii) in systems with a singular continuous spectrum the scaling with the
system size may be anomalous 
\be
\label{spe} 
\Delta \propto N^{-d/d_e}
 \ee
($d_e \neq
1$). A precise
definition of $d_e$ may depend on the system in question. In the
Harper model \cite{harper},  $d_e \approx 1/2$ stands for the
Hausdorff dimension of the spectrum.

 We are now ready to define
the dimensionless conductance in quantum chaos. In cases i) and ii)
above, $g \to \infty$ (metal) and $g = 0$ (insulator) respectively.
In case iii) (including $d_e =1$), 
 we define, with the help of Eq. ({\ref{p2}) and (\ref{spe}}),  the
dimensionless conductance in quantum chaos, \be \label{g} g(N) =
\frac{E_c}{\Delta} = N^{\gamma_{clas}} ~~~~~ \gamma_{clas} =
\frac{d}{d_e} - \frac{2}{\beta}. \ee\\

In disordered systems it is intuitively clear that if the strength
of disorder is sufficiently strong the insulator limit will be
eventually reached. In quantum chaos it is not that evident as it
may occur that the coupling constant of the model which plays the
role of disorder never reaches this limit. This is a fact to take
into account before making any statement about the existence of a
metal-insulator transition in quantum chaos.

In order to study the change of the dimensionless conductance with
the system size due to quantum effects, it is helpful to define
the function $\beta(g)$: \be \label{beta} \beta(g) =
\frac{\partial \log g(L)}{\partial \log L}. \ee If quantum effects
are not important, \be \beta(g) = \gamma_{clas} =  \frac{d}{d_e} -
\frac{2}{\beta}. \ee
Based on the above expressions and under the
assumptions of the one parameter scaling theory we propose the
following alternative
definition of  universality class in quantum chaos:\\

For systems such that $\gamma_{clas} >  0$, eigenfunctions are
delocalized as in a metal and the spectral correlations are
universally described by WD statistics.\\

For systems such that  $\gamma_{clas} < 0$, eigenfunctions are
localized as in a insulator and  the spectral correlations are
universally described by Poisson statistics.\\

For systems such that $\gamma_{clas} =  0$, eigenfunctions are
multifractal as at the metal insulator transition and the
spectral correlations are universally described by critical
statistics \cite{kravtsov}.\\

Assuming for the moment that the spectrum is not fractal, $d_e
=1$, the one parameter scaling theory predicts that if the
classical motion is such that $\beta < 2/d$  the quantum
properties are those of an insulator no matter whether the
classical dynamics is chaotic or not. For instance, for $d=3$, any
classical motion leading to $\beta < 2/3$, even if chaotic, will
drive the quantum system to the insulator phase.

Another example is the 1$d$ kicked rotor with a smooth potential
\cite{kick}, classically $\langle p^2 \rangle \propto t \to
\gamma_{clas} = -1 <0$.  The one parameter scaling  predicts that
the spectral correlations are described by Poisson statistics as
in an insulator even though the classical dynamics is chaotic. It
is remarkable that by using scaling arguments dynamical
localization can be predicted without having to map the
problem onto a 1$d$ Anderson model.  \\

Equation (\ref{beta}) also suggests that a metal insulator
transition $\gamma_{clas} = 0$ is possible in less than two
dimensions. In 1$d$, it is related to a superdiffusive classical
motion such that $\beta = 2/d_e \geq 2$. For instance in non
conservative 1$d$ systems such as the kicked rotor ($d_e = 1$),
the spectral correlations will be metal-like for $\beta > 2$,
similar to those of a metal-insulator transition for $\beta =2$
and, insulator-like for $\beta < 2$ . Therefore dynamical
localization can be overcome, even in one dimension, if the
classical diffusion is fast enough $\beta \geq  2$.

We note that these conclusions are still restricted to the
semiclassical approximation. Quantum effects such as destructive
interference may slow down or even stop the quantum motion.
However the one parameter scaling theory, as is the case of
disordered systems, is capable to handle this situation as well.
The way to proceed is to compute the Thouless energy not with the
classical but with the quantum travel time to cross the sample.
Assuming $\langle p^2 \rangle_{quan} \propto t^{\beta_q}$,  WD
(Poisson) statistics applies if $\gamma_q  = {d}/{d_e} -
{2}/{\beta_q} > (<) 0$. As an example we discuss the Harper model
\cite{harper}.  It is well known that this model undergoes a metal
insulator transition for a specific value of the coupling constant
($\lambda = 2$). We note that classically the system is
integrable. The semiclassical dimensionless conductance cannot
predict the transition. However the quantum motion $\langle x^2
\rangle \sim t^{2d_H}$ at the metal-insulator transition is
diffusive where $d_H \sim 0.5$ is the Hausdorff spectral
dimension. According to our definition Eq. (\ref{g}),  the
``quantum" dimensionless conductance $g=g_c$ is a constant
independent of the system size.  This is a signature of a metal
insulator transition. Therefore our simple method predicts
correctly the metal insulator transition in this model as well.\\

%%%%%%%%%%%%%Fig 3%%%%%%%%%%%%%%%%%%%%%%%%%%%%%%%%%%%%%%%%%%%%%%%%%%%%%%%%%%%%%%%%%%%
\begin{figure}
\includegraphics[width=0.65\columnwidth,clip]{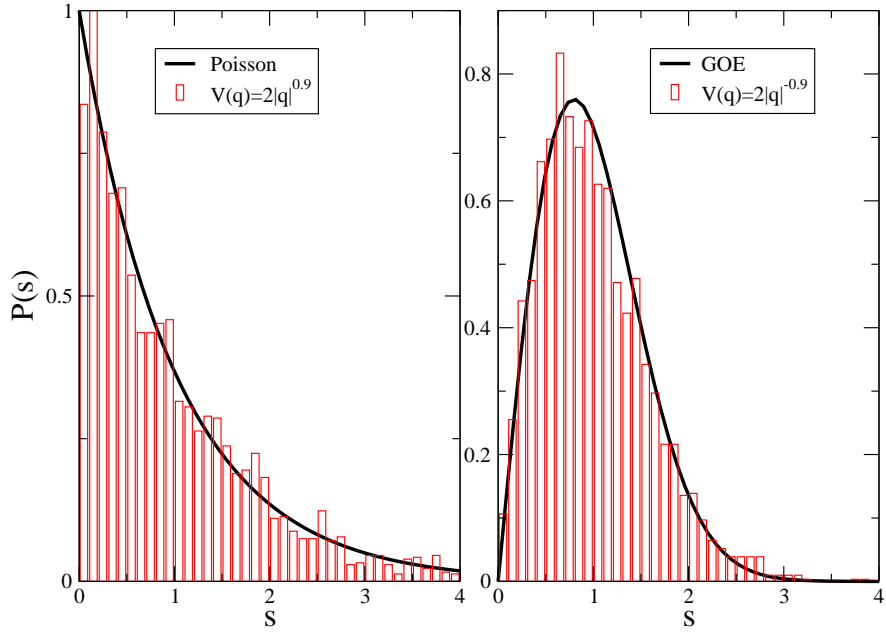}
\caption{The level spacing distribution $P(s)$ in units of the
mean level spacing $s=(E_i-E_{i-1})/\Delta$. A transition is
observed from Poisson to WD statistics (keeping $\epsilon$
constant) as $\alpha$ goes from negative to positive. In both
cases $N=3100$.} \label{figure3}
\end{figure}
%%%%%%%%%%%%%Fig 3%%%%%%%%%%%%%%%%%%%%%%%%%%%%%%%%%%%%%%%%%%%%%%%%%%%%%%%%%%%%%%%%%%%
%%%%%%%%%%%%%Fig 4%%%%%%%%%%%%%%%%%%%%%%%%%%%%%%%%%%%%%%%%%%%%%%%%%%%%%%%%%%%%%%%%%%%
\begin{figure}
\includegraphics[width=0.65\columnwidth,clip]{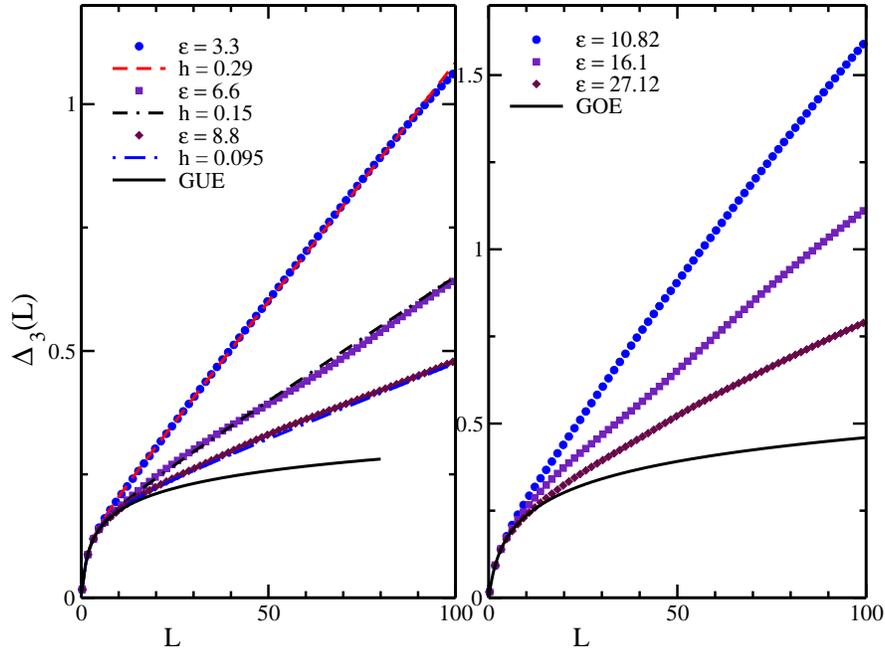}
\caption{The spectral rigidity $\Delta_3(L)$ versus $L$ for
$V(q)=\epsilon \log (q)$. The right (left) panel corresponds to
the case of (broken) time reversal invariance. The linear behavior
for $ L \gg1$ is a signature of the Anderson transition. The
slashed lines in the case of broken time reversal invariance are
the analytical prediction $\epsilon \approx 1/h$ for this type of
classical anomalous diffusion.} \label{figure4}
\end{figure}
%%%%%%%%%%%%%Fig 4%%%%%%%%%%%%%%%%%%%%%%%%%%%%%%%%%%%%%%%%%%%%%%%%%%%%%%%%%%%%%%%%%%%

\subsection{ Stability of the semiclassical predictions}

From a more technical point of view the problem of the stability
of the semiclassical predictions under quantum effects boils down
to the way in which the classical diffusion constant included in
the definition of the dimensionless conductance gets renormalized
by quantum corrections. For normal diffusion it is well known that
quantum corrections (interference) may modify substantially the
semiclassical results. For instance in a 2$d$ disordered system
\cite{mirlin,one}, a simple calculation of the one loop
perturbative quantum correction shows that the  diffusion constant
is smaller than in the classical case,  $\beta(g)$ becomes
negative and the semiclassical metal-insulator transition is
destroyed by
quantum corrections. \\

However in the case of classical anomalous diffusion 
we shall show that in certain circumstances
semiclassical predictions are stable under quantum corrections.
For the sake of simplicity we focus on the case $d=1$ and $d_e=1$.
Extension of the results to other dimensions is straightforward.
Our starting point is a classical diffusion process described by the 
following fractional Fokker-Planck equation, 
\be
\left(\frac{\partial}{\partial t}- D_{cla}\frac{\partial^{2/\beta}}
{\partial |p|^{2/\beta}}\right)P(p,t)= \delta(p)\delta(t) \nonumber
\ee
(see \cite{rev} for the definition of fractional derivative) 
where $D_{cla}$ is the classical diffusion coefficient and $\beta$ is a real number.
The moments of the distribution $P(p,t)$ (if well defined)   are given by $\langle |p|^k
\rangle \propto t^{\beta k/2}$, with $k$ a real numbers. 
In analogy with the case of normal diffusion the propagator is 
given by  $K_0(q,\omega=0)= \frac{\nu}{D_{cla}q^{{{2/\beta}}}} $
with $\nu$ the spectral density. 

The one-loop correction to the classical
propagator due to interference effects takes the following form \cite{mirlin,ever},
\begin{equation}
\label{a1}
K^{-1}(q)=K_0^{-1}(q)-\frac{(\pi\nu)^2}{2}
\int(dk)\frac{|q+k|^{{{{2/\beta}}}}-|k|^{{{{2/\beta}}}}}
{|k|^{{{{2/\beta}}}}}    \;.
\label{a4}
\end{equation}

The integral over $k$ is carried out,
with the help of the following expansion \cite{mirlin},
\begin{equation}
\label{a2} {|{ q}+{k}|^{{{2/\beta}}}\over|{
k}|^{{{2/\beta}}}}-1\simeq\left\{
\begin{array}{ll}
\displaystyle{ {{{2 \over \beta}}}{{ qk}\over k^2}+{{{{1}}}\over
\beta}{q^2\over k^2}+ {{{2 \over \beta}}}\left({{{{1}}}\over
\beta}-1\right)\left({{\bf qk}\over k^2}\right)^2 +\ldots   }
\ ,&\qquad q\ll k\\
\displaystyle{ {|q|^{{{2/\beta}}}\over |k|^{{{2/\beta}}}}     } \
,&\qquad q\gg k
\end{array}\right.  \;.
\end{equation}
where $L$, the system size, acts as an infrared cutoff.

The quantum diffusion constant $D_{qua}$ to one loop is easily
obtained from Eq.({\ref{a2}),(\ref{a1}), \be
{D_{qua}}&=&D_{cla}-\mbox{C}  L^{{{{2/\beta}}}-1}   ~~~~ 1 < {{{2/\beta}}} < 2\\
{D_{qua}}&=&D_{cla}-\mbox{C}  \log(qL)   ~~~~ {{{2/\beta}}} = 1\\
\nonumber {D_{qua}}&=&D_{cla}-\mbox{C}  q^{1-{{{2/\beta}}}}  ~~~~ 0
<{{{2/\beta}}} < 1\\  \label{a3} \ee where $C$ is a
different constant for each case.

The importance of the quantum effects depends strongly on the
value of ${{{2/\beta}}}$. In the region $1 < {{{2/\beta}}} < 2$
quantum corrections diminish the value of the classical diffusion
constant. These corrections, due to its dependence on the system
size, grow as higher order in perturbation theory are taken into
account and eventually induce a transition to localization
\cite{mirlin}.\\

In the region $0<{{{2/\beta}}} <1$ the correction is of higher
order in $q$ than the classical term. This feature is not modified
by higher orders in perturbation theory \cite{mirlin} and
consequently
the classical diffusion is not normalized.  \\

The case ${{{2/\beta}}} =1$, related to the metal-insulator
transition, deserves a special consideration. The logarithmic
behavior resembles superficially that of 2$d$ disordered system,
\begin{equation}
\label{2d}
D_{qua}=D_{cla}-\mbox C\ln(L/l)\ ,
\end{equation}

However there are essential differences. Comparing the two
formulas, we see that in the case of anomalous diffusion the mean
free path $l$ in the 2$d$ case is replaced by the inverse momentum
$q^{-1}$. Therefore, the correction to the bare coupling constant, unlike the $2D$ disordered 
conductor,
is small for low momenta $q\sim 1/L$.  Qualitatively this implies the {\it
absence} of eigenstate localization.  For a rigorous proof based on the 
evaluation of $\beta(g)$ including higher order terms we refer to \cite{mirlin} and reference therein.

Therefore a  Hamiltonian with
classical dynamics such that ${{{2/\beta}}} =1$ should have
quantum properties similar to that of a disordered conductor at
the metal-insulator transition.  If the classical diffusion is
slower (${{{2/\beta}}} < 1$)  quantum corrections will eventually
induce a transition to localization. A final remark is in order.
We recall that by dimension we do not mean the spatial dimension
but the effective dimension in momentum space.  A  conservative
one dimensional  (in space) system is always integrable so no
transition to delocalization can be observed.  However in momentum
space the effective dimension is zero, just a point defined by
energy conservation. But if energy is not conserved, as is the
case in kicked rotors, the effective dimension in momentum space
is the same that the spatial one, namely, the unity.\\

To summarize: We have identified three universality class in
quantum chaos: metal, insulator and  metal-insulator transition.
According to the one parameter scaling theory these are the only
windows of universality. If the classical diffusion is anomalous
we have shown in what  circumstances semiclassical predictions are
not substantially modified by quantum effects.  Two ways to reach
the metal insulator transition in quantum chaos are identified: By
classical anomalous diffusion such that $\gamma_{clas} = 0 $ or by
quantum effects $\gamma_q =0$ in a classical Hamiltonian with
normal diffusion and $\gamma_{clas} > 0$.

\section{Scaling and  the Anderson transition in quantum chaos}
We now test the predictions of the previous section in two
specific models: A 3$d$ kicked rotor and a 1$d$ kicked rotor with
a classical singularity. In the former the classical diffusion is
normal.  In the latter it is anomalous with an exponent which
depends on the type of singularity. We shall show that by tuning
the kicking strength (in 3$d$) or the type of singularity (1$d$)
to a specific value we can induce a metal insulator transition in
quantum chaos.  However the physical origin is slightly different
in both situations. In 3$d$ it is caused by quantum destructive
interference effects that slow down the classical chaotic motion
while in 1$d$ has its origin in the underlying classical anomalous
diffusion. Finally we mention other systems that, according to the
classification scheme introduced in the previous section also
belong to the metal insulator universality class.

\subsection{Anderson transition induced by classical anomalous diffusion}
Previously we determine by using the one parameter scaling theory that classical superdiffusion may induce a metal-insulator transition even in 1d. 
In this section we provide numerical support to this theoretical prediction.   In order to proceed we need to 
choose a system in which classical diffusion is anomalous.  A suitable candidate is the kicked rotor with a non analytical potential 
introduced in \cite{ant9},
 \be
\label{ourmodel} {\cal H}= \frac{p^2}2 +V(q)\sum_n\delta(t -nT)
\ee with $V(q)=\epsilon |q|^\alpha$ and $V(q)=\epsilon\log(|q|)$;
$q\in [-\pi,\pi)$, $\alpha \in [-1,1]$ and $\epsilon$ a real
number (for other type of singularities see
\cite{bao,liu,antwan2}). The dynamical evolution over $T$ is
dictated by the map:
%following discrete version of Hamilton's equations
$p_{n+1}=p_{n}- \frac{\partial V(q_n)}{\partial q_n}$, $q_{n+1}=q_n+Tp_{n+1}$
(mod$~2\pi$).
%We focus on the statistical properties of
%the dynamics in momentum space.
%%%%%%%%%%%%%Fig 5%%%%%%%%%%%%%%%%%%%%%%%%%%%%%%%%%%%%%%%%%%%%%%%%%%%%%%%%%%%%%%%%%%%
\begin{figure*}
\includegraphics[width=0.45\columnwidth,clip]{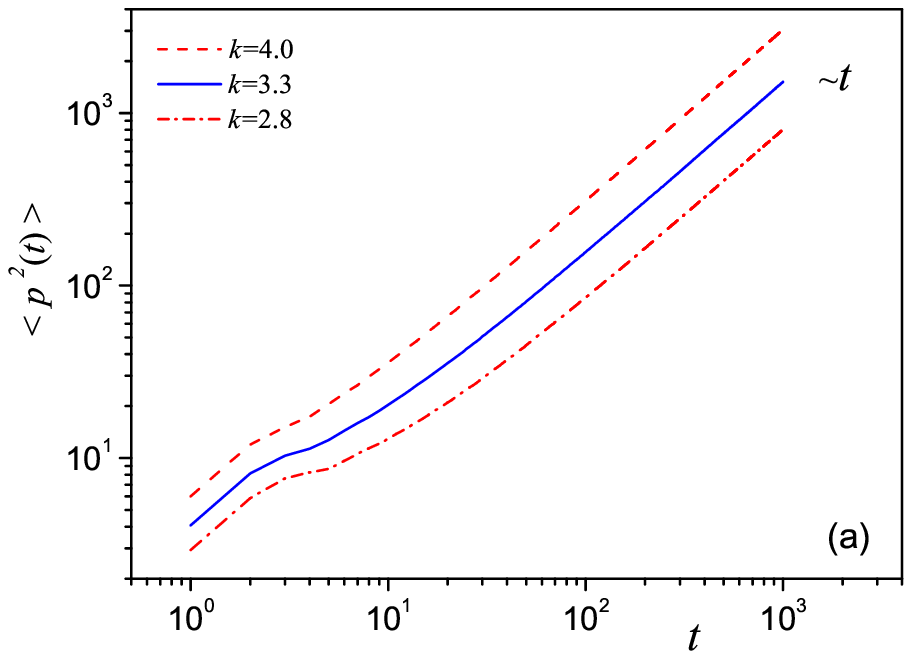}
\includegraphics[width=0.45\columnwidth,clip]{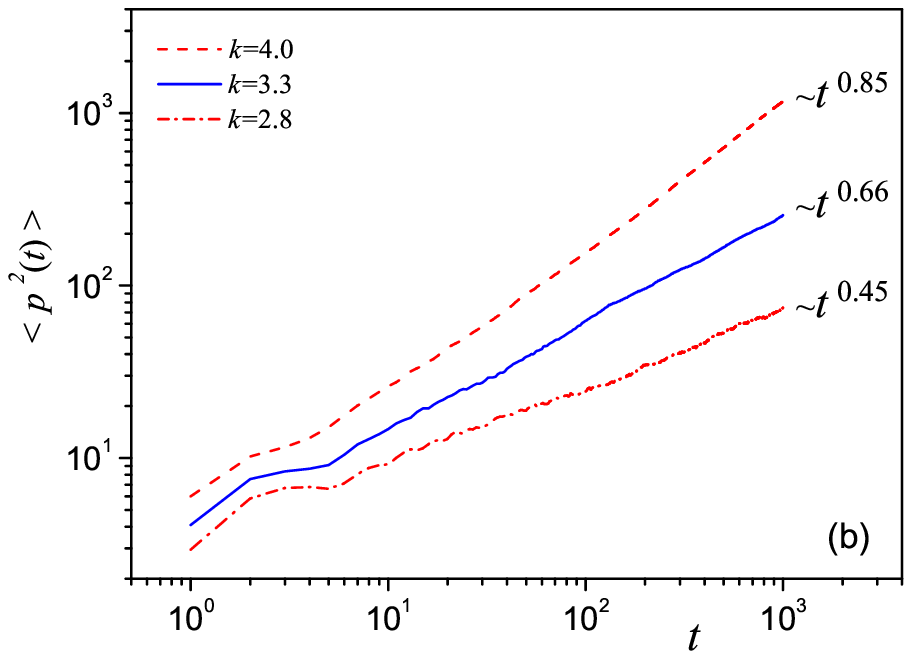}
\caption{Classical and quantum $\langle p^2 \rangle$ for three
different kicking values, above, below and at the transition $k =
k_c \approx 3.3$ in  the 3$d$ kicked rotor with a smooth
potential. Classical diffusion is normal however quantum
destructive interference slow down and eventually arrest the
diffusion for $k < k_c$.  In agreement with the one parameter
scaling theory, at $k = k_c$, $\langle p^2 \rangle \propto
t^{2/3}$.} \label{figure5}
\end{figure*}
%%%%%%%%%%%%%Fig 5%%%%%%%%%%%%%%%%%%%%%%%%%%%%%%%%%%%%%%%%%%%%%%%%%%%%%%%%%%%%%%%%%%%

%%%%%%%%%%%%%Fig 6%%%%%%%%%%%%%%%%%%%%%%%%%%%%%%%%%%%%%%%%%%%%%%%%%%%%%%%%%%%%%%%%%%%
\begin{figure*}
\includegraphics[width=0.45\columnwidth,clip]{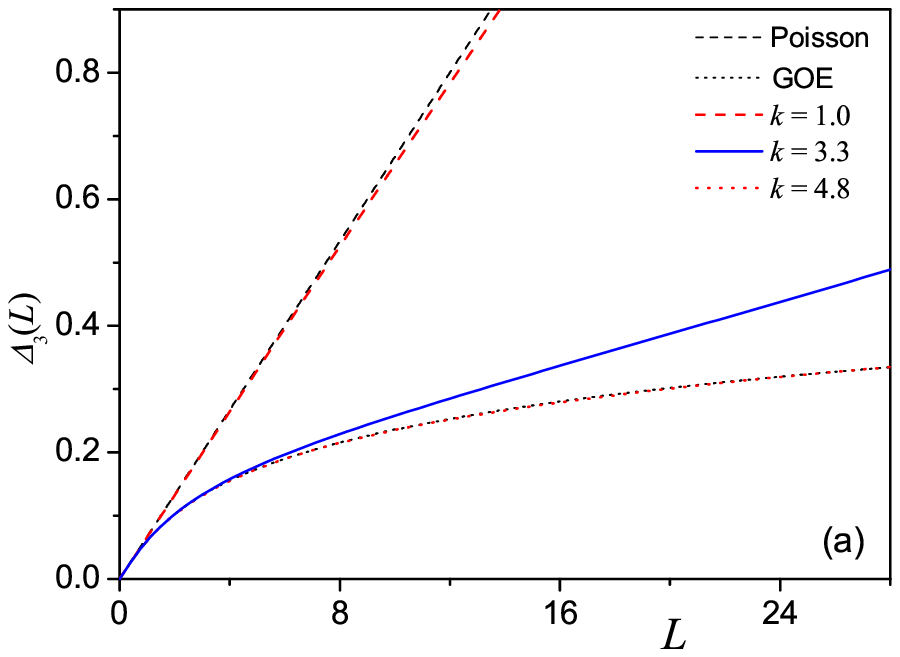}
\includegraphics[width=0.45\columnwidth,clip]{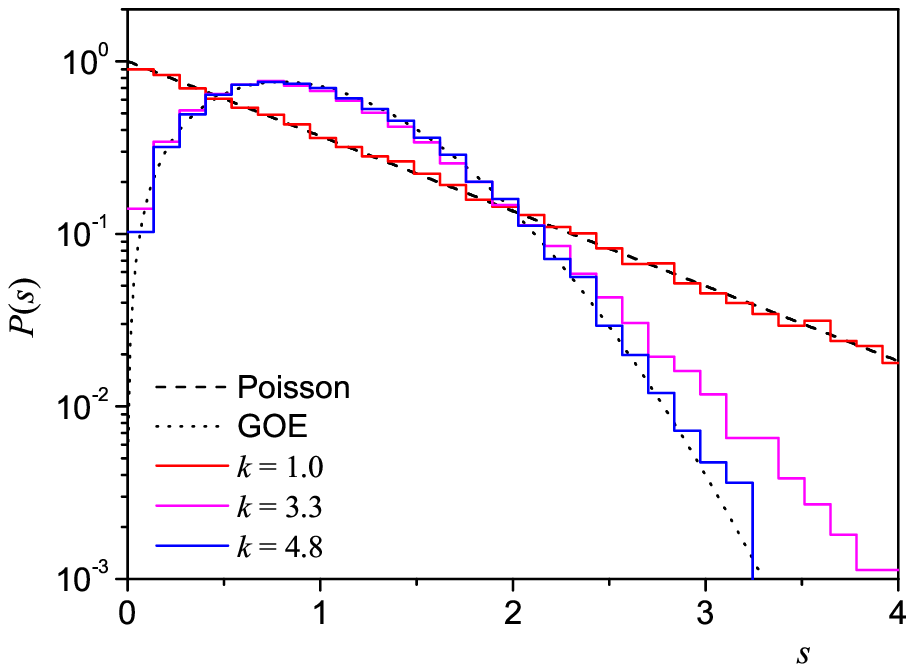}
\caption{The spectral rigidity $\Delta_3(L)$ versus $L$ and level
spacing $P(s)$ for three different kicking values, above, below
and at the transition $k = k_c \approx 3.3$ in  the 3$d$ kicked
rotor with a smooth potential.  A transition from WD to Poisson
statistics is clearly observed as the kicking strength crosses the
critical value $k \approx k_c$ from above.  For $k \sim k_c$ the
level statistics have all the signatures of an metal-insulator
transition such as linear $\Delta_3(L)$, level repulsion for $s
\to 0$ as in a metal but exponential decay of $P(s)$ for $s
>1$ as  in an insulator.  Although it is not shown the spectrum is
scale invariant at $k \approx k_c$.} \label{figure6}
\end{figure*}
%%%%%%%%%%%%%Fig 6%%%%%%%%%%%%%%%%%%%%%%%%%%%%%%%%%%%%%%%%%%%%%%%%%%%%%%%%%%%%%%%%%%%

We have evaluated the classical density of probability $P(p,t)$
and its moments $\langle p^k \rangle$ by evolving $10^8$ initial
conditions uniformly distributed in $(q,p)\in[-\pi,\pi)\times
[-\pi,\pi)$.  For $\alpha> 0.5$ (see Fig. 1), the diffusion is
normal and $P(p,t) \sim e^{-cp^2/t}/\sqrt{t}$ ($c$ a constant).
However, for $-0.5<\alpha < 0.5$ and $t > p$, $P(p,t) \sim t
p^{\gamma(\alpha)}$ has power-law tails (see inset in Fig. 1 for a
relation between $\gamma$ and $\alpha$). Such tails are considered 
a signature of anomalous diffusion. As a consequence, the
classical moments of the distribution scale as $\langle p^k
\rangle \sim t^{(1-\alpha)k}$ (see Fig. 1) and the dimensionless
conductance is given by \be \beta(g) = -\frac{\alpha}{1-\alpha}. \ee According to
the findings of the previous section, for $\alpha \leq 0$ quantum
corrections do not modify qualitatively the semiclassical
prediction for $\beta(g)$. Therefore according to the one
parameter scaling theory we expect WD statistics for $\alpha < 0$,
Poisson statistics for $\alpha > 0$ (quantum corrections will only
make the transition to localization faster) and an Anderson
transition for $\alpha = 0$
that corresponds naturally with the case of a $\log$ singularity.\\

We now compare these theoretical predictions with numerical
results. In order to proceed we evaluate the quantum evolution
operator $\cal U$ over a period $T$.  After a period $T$, an
initial state $\psi_0$ evolves to $\psi(T) = {\cal U}\psi_0 =
e^{\frac{-i {\hat p}^2T}{4{\bar h}}} e^{-\frac{iV(\hat q)}{\bar
h}}e^{\frac{-i {\hat p}^2 T}{4{\bar h}}}\psi_0$ where $\hat p$ and
$\hat q$ stand for the usual momentum and position operator. The
spectrum was obtained by solving the eigenvalue problem ${\cal
U}\Psi_{n}=e^{-i\kappa_n/ \hbar}\Psi_{n}$ where $\Psi_{n}$ is an
eigenstate of $\cal U$ with eigenvalue $\kappa_n$.  We express the
evolution operator in a matrix form  $\langle m| {\cal U} | n
\rangle $ in the basis of the momentum eigenstates $\{| n \rangle
= \frac {e^{in \theta}}{\sqrt{2\pi}}\}$.
%by using a basis of plane
%waves $| n \rangle = \frac{e^{in \theta}}{\sqrt{2\pi}}$
%(where $n,m=1,\ldots N$)
%which are the eigenvalues of the momentum operator.
%The resulting matrix  is Unitary
%exclusively in the thermodynamic limit.
The resulting evolution matrix
(for $N$ odd) then reads
\be
\label{uni}
\langle m| {\cal U}| n \rangle = \frac{1}{N}e^{-i2\pi M n^2/N}
\sum_{l}e^{i\phi(l,m,n)}
\ee
where $\phi(l,m,n)= 2\pi (l+\theta_0)(m-n)/N-iV(2\pi (l+\theta_0)/N)$,
%$m,n = 1,\ldots N$,
$l = -(N-1)/2,\ldots (N-1)/2$ and $0 \le \theta_0 \le 1$; $\theta_0 $
is a parameter depending on the boundary conditions ($\theta_0=0$ for
periodic boundary conditions). The eigenvalues and eigenvectors of $\cal U$
can now be computed by using standard diagonalization techniques.\\

We first analyze the  moments of the quantum density of of
probability, $\langle p^k \rangle$. In Fig. 2 (right) it is
observed that in agreement with the perturbative analysis above
the classical time dependence is not modified by quantum
corrections. However the diffusion constant is smaller in the
quantum case due to quantum destructive interference effects (Fig.
2 left). Similar results are obtained for $\alpha < 0$. For
$\alpha
>0$ quantum corrections can only make the transition to
localization faster.  In conclusion,  for $\alpha \leq 0$ the
scaling of the dimensionless conductance does not change with
respect to the semiclassical limit $g \propto L^{-\alpha}$ so
a genuine metal-insulator transition is expected at $\alpha = 0$.\\

The analysis of the level  statistics confirms this theoretical
prediction. The level spacing distribution $P(s)$ (see Fig. 3)
tends to Poisson (WD) for $\alpha > 0$ ($\alpha < 0$).  As
$\alpha$ approaches zero it  is much more difficult to observe the
transition to either WD or Poisson statistics. The reason for that
is simple.  The dimensionless conductance goes as $g(L) \propto
L^{-\alpha}$ but the universal WD (Poisson) limit are only
expected for $g \to \infty$ ($g \to 0$). Therefore, as we approach
the transition $\alpha =0$, the universal limits are reached very
slowly.\\

We now study the case of a potential with log singularity  which
according to our prediction it is related to the Anderson
transition. In order to proceed we also evaluate long range
correlators as the spectral rigidity
$\Delta_{3}(L)=\frac{2}{L^4}\int_{0}^{L}(L^3-2L^2x+x^3)\Sigma/2^{2}(x)dx$
where $\Sigma^{2}(L)=\langle L^2 \rangle - \langle L \rangle^2 = L
+2\int_{0}^{L} (L-s) R_{2}(s)ds$ is the number variance and
$R_2(s)$ is the two level spectral correlations function. The
$R_2(s)$ associated to this type of motion has been evaluated in
Ref.\cite{ant7} for systems with broken time reversal invariance,
\be \label{cri} R_2(s)=-K^2(s)=-\frac{\pi^2
h^2}{4}\frac{\sin^2(\pi s)}{\sinh^2(\pi^2 h s/2)} \ee where $h \ll
1$ is related to $\epsilon$ by $h =1/\epsilon \ll 1$. In Fig. 4 we
show $\Delta_3(L)$ for different $\epsilon$'s and $N=3099$. The
time reversal invariance in Eq.(\ref{uni}) is broken by setting
$T=2\pi \beta$ with $\beta$ an irrational number. As observed in
Fig. 4 (left), $\Delta_3(L)$ is asymptotically linear, this
feature is typical of a disordered conductor at the
metal-insulator transition. Moreover, the agreement with the
analytical prediction based on Eq. (\ref{cri}) is excellent. We
remark that the value of $h$ best fitting the numerical result is
within $5 \%$ of the analytical estimate $h = 1/\epsilon$. We have
repeated the calculation keeping the time reversal invariance
($T=2\pi M/N$) of Eq.(\ref{cri}) (Fig. 4, right). We do not have
in this case an analytical result to compare with, but
$\Delta_3(L)$ is also asymptotically linear as in the previous
case.

These numerical results clearly support  the analytical results of
the previous section. Three and only three universality class are
observed: metal, insulator and metal-insulator transition. Finally
we note that the novel semiclassical characterization of the
metal-insulator transition proposed in this paper may be of
interest for experimental studies about Anderson localization, as
it enlarges the number of physical systems in which this
phenomenon can be studied.

\subsection{Anderson transition induced by quantum fluctuations}

We now study the other route to the Anderson transition, namely,
through quantum destructive interference effects that slow down an
otherwise classically chaotic motion.

Our starting point is again the kicked rotor introduced in the
previous section. However this time it is defined in 3$d$ and with
a smooth potential $V(\theta_1,\theta_2,
\theta_3)=k\cos(\theta_1)\cos(\theta_2)\cos(\theta_3)$, the
classical diffusion in momentum space is thus expected to be
normal provided that the classical space is fully chaotic. We note
if the spectrum is not fractal ($d_e =1$) and the diffusion is
normal, the dimensionality must be larger than two in order to
observe a metal insulator transition.\\

In Fig. 5 (left) it is shown that as was expected classical motion
is diffusive $\langle p^2 \rangle \propto t$ in the range of $k$
of interest. Quantum dynamics depends strongly on $k$. In analogy
with a 3$d$ disordered system we expect  that strong Anderson
localization effects stop the classical diffusion for sufficiently
small $k$. In the opposite limit quantum effect should not
modified substantially the classical motion.  We carry out
\cite{antwan7} a careful finite size scaling analysis in order to
determine how localization corrections depend on $k$. A
metal-insulator transition of purely quantum origin is located at
$k =k_c \sim 3.3$.  Quantum diffusion becomes anomalous with
$\beta =2/3$ (see Fig. 5 (right)) in agreement with the prediction
of the one parameter scaling theory. In consequence the level
statistics  should be described by WD (Poisson)  statistics in the
limits $k  \gg (\ll) k_c$.

As is shown in Fig. 6, the numerical results fully agree with
these theoretical predictions.  For $k =k_c$ the spectral
correlations have all the features of an 3$d$ Anderson transition
including the numerical value of parameters such as the slope of
the number variance $\sim 0.2$.

The kicked rotors studied in this talk are not the only ones that
belong to the universality class of the metal-insulator
transition. Other models that falls in it, at
least for some value of the coupling constant,  include, Harper
model \cite{harper}, quantum exchange interval maps \cite{bogo04},
Coulomb billiards \cite{altshu}, the Kepler problem \cite{wintgen} and generalized kicked rotors \cite{bao,liu}.

\section{Conclusions}
The main findings of this paper can be summarized as follows: 
 For the first time we have adapted the one parameter scaling theory
to the context of quantum chaos.  Then we have
utilized it to determine the number of universality class in
quantum chaos and propose a more accurate definition of them.  The 
one parameter scaling theory has permitted us to give for the first time a precise definition
of the 
universality class related to the metal insulator transition. We have identified two routes
(semiclassical and purely quantum) to reach a metal-insulator
transition in quantum chaos.  In order to support our theoretical
conclusions we have investigated numerically a 1d kicked rotor
with singularities where the Anderson transition has a
semiclassical origin related to classical superdiffusion and the 3d kicked rotor with a smooth
potential where the transition to localization is induced by
destructive interference effects. Our findings open the
possibility of studying the metal insulator transition
experimentally in a much broader type of systems.\\

AMG acknowledges financial support from a Marie Curie
Outgoing Action, contract MOIF-CT-2005-007300.  JW is grateful to professor C.-H. Lai for his encouragement and
support, and acknowledges support from Defence Science and Technology
Agency (DSTA) of Singapore under agreement of POD0613356.

\end{document}